\documentclass[pre,aps,amsfonts,amssymb,amsmath,graphicx,showpacs,twocolumn,epsfig,amscd,ulem]{revtex4}

\usepackage{graphicx}
\usepackage{natbib}
\usepackage{amssymb}
\usepackage{amsmath}
\usepackage{verbatim}

\begin{document}

\title{High-bandwidth viscoelastic properties of aging colloidal glasses and gels}

\author{S.\ Jabbari-Farouji$^{1,2}$, M.\ Atakhorram$^3$,D.\ Mizuno$^{3,4}$, E.\ Eiser$^{5,6}$, G.H.\ Wegdam$^1$,  F.C.\ MacKintosh$^3$,  Daniel Bonn$^{1,7}$, C.F.\ Schmidt$^{4,8}$}

\affiliation{$^1$van der Waals-Zeeman Institut, Universiteit van
Amsterdam, 1018XE Amsterdam, The Netherlands\\
$^2$ Group Polymer Physics, Department of Applied Physics, Technische Universiteit Eindhoven
 5600MB Eindhoven, The Netherlands\\
  $^3$Divisie Natuur- en Sterrenkunde, Vrije Universiteit Amsterdam, 1081HV Amsterdam, The Netherlands\\
 $^4$ Organization for the Promotion of Advanced Research, Kyushu University, Higashi-ku, Hakozaki 6-10-1, 812-0054 Fukuoka,
Japan\\
$^5$van 't Hoff Institute for Molecular Sciences, Universiteit van Amsterdam, 1018WV Amsterdam, The Netherlands\\
$^6$University of Cambridge, Department of Physics,Cavendish Laboratory,J J Thomson Avenue, Cambridge CB3 0HE, UK\\
$^7$Laboratoire de Physique Statistique de l'ENS, 75231 Paris Cedex 05, France\\
$^8$ Physikalisches Institut, Georg-August-Universität, 37077 G\"{o}ttingen, Germany}
\date{\today}

\begin{abstract}
We report measurements of the frequency-dependent shear moduli of
aging colloidal systems that evolve from a purely low-viscosity
liquid to a predominantly elastic glass or gel. Using
microrheology, we measure the local complex shear modulus
$G^{*}(\omega)$ over a very wide range of frequencies (1 Hz- 100
kHz). The combined use of one- and two-particle microrheology allows us to differentiate between colloidal glasses and gels - the glass is homogenous,
whereas the colloidal gel shows a considerable degree of
heterogeneity on length scales larger than 0.5 micrometer.
Despite this characteristic difference, both systems exhibit similar
rheological behavior which evolve in time with aging, showing a crossover
from a single power-law frequency dependence of the viscoelastic modulus
to a sum of two power laws. The crossover occurs at a
time $t_{0}$, which defines a mechanical transition point. We
found that the data acquired during the aging of different samples
can be collapsed onto a single master curve by scaling the aging
time with $t_{0}$. This raises questions about the prior interpretation
of two power laws in terms of a superposition of an elastic network
embedded in a viscoelastic background.
 {\it keywords:  Aging, colloidal glass, passive microrheology }
\end{abstract}

\maketitle

\section{Introduction}

Soft glassy materials are ubiquitous in everyday life. A
common feature of all such materials is their relatively large
response to small forces (hence soft) and their disordered (glassy) nature. Pertinent examples of such systems are foams, gels,
slurries, concentrated polymer solutions and colloidal
suspensions. These systems show interesting viscoelastic
properties; depending on the frequency with which they are
perturbed, they can behave either liquid- or solid-like. In
spite of their importance for numerous applications, the
mechanical behavior of such soft glassy materials is still
incompletely understood \cite{Sollich}.

In recent decades, colloidal suspensions have been used extensively as model
systems for the glass transition in simple liquids \cite{Megen,kroon,glass}
and gel formation \cite{gel1}; since the diffusion of the particles can easily
be measured using, e.g., light scattering or confocal microscopy \cite{Kegel}.
The viscoelasticity of such systems, especially its development during the aging of glassy systems or the formation of a gel, has, however, received relatively little attention.

\begin{figure}
\begin{center}
\includegraphics[scale=0.45]{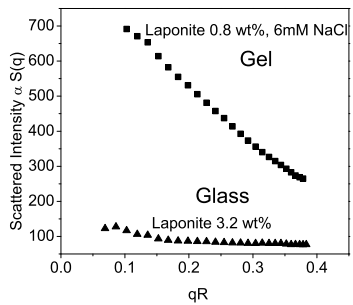}
\end{center}
 \caption{Light scattering intensity relative reduced with respect to scattered intensity from toluene for two samples
 a)Laponite 3.2 wt\% and b) Laponite 0.8 wt\%, 6 mM NaCl. The data
 points are taken at late stages of aging when the scattering intensity has
 stabilized.
 }\label{fig1}
\end{figure}

Another issue that deserves attention is differentiating between colloidal gels
and glasses in terms of their rheological properties \cite{Pham2006}.
The main difference between colloidal gels and glasses
stems from their structure. While the glass has a homogenous
liquid-like structure with no long-range order, a gel can have a
heterogeneous structure whose characteristic length is set by the
mesh size of the gel network (see Fig. \ref{fig1}) \cite{PRL}.

As an example of a soft glassy system, we here focus on the viscoelasticity of Laponite
suspensions for which a very rich phase diagram has been reported \cite{Mourchid2000,Italian}.
When dissolved in water, Laponite suspensions evolve from a liquid-like state to a non-ergodic
solid-like state \cite{kroon,Bonn2,glass,Bellour}. During this process
the mobility of the particles slows down and viscoelasticity develops.
This system is an interesting one to study since both colloidal
gels and glasses can be obtained depending on Laponite
concentration and added salt content \cite{PRL}. Therefore, it
provides us with the possibility to investigate the similarities
and differences in the viscoelastic properties of the two types of non-equilibrium states.

To study the mechanical properties of gels and glasses, we used
microrheology (MR), which allows us to measure frequency-dependent
shear moduli over a wide range of frequencies. This technique is based on the
detection of small displacements of probe particles embedded in
the soft glassy material, from which we obtain the mechanical
properties of surrounding matrix \cite{Gittes2}. Considering the fragility of
soft materials, this technique is ideally suited for our studies, since it is
less invasive than conventional rheometry.

Here we have used a combination of one and two-particle MR
Measurements \cite{Maryam,Maryam2} to probe the mechanical properties and possible inhomogeneities of colloidal
gels and glasses on length scales of the order of the particle
size $1 \mu$m and separation distances of the order of 5-20
$\mu$m.

\section{Experimental}

\subsection{Materials}

We have studied charged colloidal disks of Laponite XLG, with an average
radius of 15 nm and a thickness of 1 nm. Laponite can absorb water,
increasing its weight up to 20\%. Therefore, we first dried it in
an oven at $100^{o}$C for one week and subsequently stored it in a
desiccator.

We prepared a number of Laponite samples with different
concentrations and salt contents. Laponite solutions without added
salt were prepared in ultra pure Millipore water (18.2 M$\Omega$
cm$^{-1})$ and were stirred vigorously with a magnet for 1.5 hours
to make sure that the Laponite particles were fully dispersed. The
dispersions were filtered using Millipore Millex AA 0.8 $\mu$m
filter units to obtain a reproducible initial state \cite{glass}.
This instant defined the zero of waiting time, $t_{w}=0$.

The Laponite solutions with pH=10 were
obtained by mixing the Laponite with a $10^{-4}$ mole/l solution of NaOH in Millipore water.
 The samples with non-zero salt content were prepared by
diluting the Laponite suspensions in pure water with a more
concentrated salt solution \cite{Nicolai2000}. For instance, a
sample of 0.8 wt \%, 6mM NaCl was prepared by mixing equal volumes
of 1.6 wt\% Laponite solution in pure water with a 12mM salt
solution.

 For the microrheology measurements, we added a small
fraction, below $ 10^{-4}$ vol \%, of silica beads with a diameter
of 1.16 $\mu$m $\pm 5 \%$ \cite{Utrecht} immediately after the
preparation of the sample. Subsequently we infused the solution into
a sample chamber of about 50 $\mu$l volume, consisting of a
coverslip and microscope slide separated by spacers of
double-sided tape with a thickness of 70 $\mu$m, sealed with
vacuum grease at the ends to avoid evaporation of the sample. All the
experiments were performed at room temperature $(21\pm 1^\circ
$C). After placing the sample chamber into the microscope, we trapped
two beads and moved them to about 20 $\mu$m above the bottom glass surface.

\subsection{Microrheology}

The experimental setup for performing one- and two-particle MR
consists of two optical tweezers formed by two
independent, polarized laser beams $\lambda_{1}=1064 $ nm (Nd:YVO4, CW)
and $\lambda_{2}=830$ nm (diode laser, CW) which can trap two
particles at a variable separation $r$. Details of the experimental setup can be found in
\cite{Maryam2,Mthesis}. Stable trapping
is achieved using a high numerical aperture objective lens which
is part of a custom-built inverted microscope. Two lenses in a
telescope configuration allow us to control of the position of the beam
foci in the plane perpendicular to the beam directions. The two
beams are focused into the sample chamber through a high numerical
objective of the microscope ($100\times$, NA 1.3).

 Back-focal-plane interferometry is used to measure the position
fluctuations of the probe bead away from the trap center
\cite{Gittes1}. The signals emerging from each of the traps are
separately projected onto two independent quadrant photodiodes,
yielding a spatial resolution for the particle position that is
better than $1$ nm.

During our measurements, the power of
each laser was typically less than 10 mW. Labview software was used to acquire
time series data of particle positions from the QPDs for a minimum
time of 45s. The data were digitized with an A/D converter at 195 kHz
sampling rate.

\subsection{Macrorheology}

The viscoelastic moduli during the aging process were also measured
using a conventional Anton Paar Physica MCR300 rheometer in
Couette geometry. To avoid perturbing the sample during the aging
process we performed the oscillatory shear measurements with a
strain amplitude of 0.01 in the frequency range of 0.1-10 Hz. In
order to prevent evaporation during the long time measurements, we
installed a vapor trap.

\subsection{Light scattering}

Our light scattering setup (ALV) is based on a He-Ne laser
($\lambda$ = 632.8 nm , 35 mW) and avalanche photodiodes as
detectors. Static light scattering experiments were performed (scattering angle range 20-150$^o$
on non-ergodic samples at late stages of aging when the scattered
intensity had stabilized. Samples were rotated to
average over different positions in the sample.

\section{Theory and data analysis}

There are two classes of microrheology (MR) techniques:
active (AMR) and passive (PMR) \cite{MR-review}. In the first of these, the
response of a probe particle to a calibrated force is measured.
In the second approach, only passive, thermal fluctuations are monitored,
from which one can infer the response function and the rheological properties of the surrounding medium using the
fluctuation-dissipation theorem (FDT). Applying the FDT assumes
thermal equilibrium. It may therefore be potentially problematic to apply PMR to
such non-equilibrium systems as aging glasses.
Nevertheless, in a prior study \cite{jabbari}, we not only directly
confirmed the validity of the FDT, but also found excellent agreement between active and passive methods in the slowly aging Laponite glass.
Therefore, in what follows we will use only the passive method, which has
the significant advantage that, with a single measurement of the
fluctuation power spectrum, one can determine the complex shear modulus simultaneously over a wide range of frequencies \cite{Weitz,Gittes2,Buchanan1,addas1}.

\subsection{One-particle Microrheology}

In one-particle MR, we first extract the complex compliance from the
position fluctuations of one particle. The time-series data of the
bead displacement measured by the quadrant photodiode is Fourier
transformed to calculate the power spectral density of displacement
fluctuations:
\begin{equation}
\label{eq:a1}  \langle|x(\omega)|^{2}\rangle= \int^\infty_{-\infty}\langle
x(t)x(0)\rangle e^{i\omega t}dt
\end{equation}
This is done for $x$ and $y$ directions in the plane normal to the laser beam. The power
spectral density of the thermal fluctuations of the probe is related to
the imaginary part of the complex compliance $\alpha(\omega)=
\alpha'(\omega)+ i \alpha''(\omega) $ via the FDT:
\begin{equation}
 \alpha ''(\omega)= \frac{\omega \langle|x(\omega
)|^{2}\rangle}{2k_{B}T}.
\end{equation}
Provided that $\alpha''(\omega)$ is known over a large enough range of
frequencies, one can recover the real part of the response function
from a Kramers-Kronig (principal value) integral:
\begin{equation}
\alpha'(\omega)=\frac{2}{\pi}P\int^\infty_0 \frac{\omega' \alpha
''(\omega')}{\omega'^{2}-\omega^{2}}d \omega'.
\end{equation}
Before calculating
the shear modulus from the response function, we calibrate the setup and correct for
the trap stiffness that shows up at low frequencies as explained
in detail in \cite{Gittes1,Maryam}.

The complex shear modulus $G^{*}(\omega)=G'(\omega)+ iG''(\omega)$ can be
obtained from the corrected complex compliance through the generalized Stokes
relation, valid for incompressible and homogenous viscoelastic materials \cite{Weitz,Gittes2}
\begin{equation}
\label{eq:a9} G^{*}(\omega)=\frac{1}{6 \pi R \alpha(\omega )},
\end{equation}
where $R$ is the radius of the probe bead.

\subsection{Two-particle microrheology}

In two-particle MR, we calculate the correlated fluctuations of
two probe beads inside the material. Such measurements
probe the viscoelastic properties of the medium on length scales
comparable to the interparticle separation. In general, with more than one probe
particle, the displacement of particle
$m$ in direction $i$ is related to the force applied to particle
$n$ in direction $j$ via the complex response tensor
$u^{(m)}_{i}(\omega)=\alpha_{ij}^{(m,n)}(\omega)F_{j}^{(n)}(\omega)$.
In the case of two particles, the response tensors
$\alpha_{ij}^{(1,1)}$ and $\alpha_{ij}^{(2,2)}$ describe how each
of the particles 1 and 2 respond to the forces applied to
the particle itself, while $\alpha_{ij}^{(1,2)}$ describes how
particle 1 responds to the forces on particle 2.

In thermal equilibrium and in the absence of external forces, the FDT
again relates the imaginary part of the response tensor to the spectrum of
displacement fluctuations of the particles.
\begin{equation}
\label{eq:a10}\alpha_{ij}^{(m,n)}(\omega)=\frac{\omega}{2k_{B}T}
S_{ij}^{(m,n)}(\omega),
\end{equation}
where the spectra of thermal fluctuations  $S_{ij}^{(m,n)}$ are
defined as
\begin{equation}
\label{eq:a11} S_{ij}^{(m,n)}(\omega)=\int_{-\infty}^{\infty} \langle
u^{(m)}_{i}(t)u^{(n)}_{j}(0) \rangle e^{i \omega t}dt.
\end{equation}
The problem of two hydrodynamically correlated particles in a
viscoelastic medium and the relation between the response tensor
and the rheological properties of the medium has been worked out
in \cite{Levine2}. The self-parts of response tensor
$\alpha_{ii}^{(1,2)}$ are the same as the ones obtained from one-particle microrheology.

The cross component part of the response tensor
$\alpha_{ij}^{(1,2)}$ can be decomposed into two parts
$\alpha_{\parallel}$ parallel to the vector $\textbf{r}$ separating the
two beads and $\alpha_{\perp}$ perpendicular to $\textbf{r}$:
$\alpha_{ij}^{(1,2)}=
\alpha_{\parallel}\hat{r_{i}}\hat{r_{j}}+\alpha_{\perp}(\delta_{ij}-\hat{r_{i}}\hat{r_{j}})$.
For incompressible fluids each of the  components are related to
the complex shear modulus via a generalization of the Oseen tensor:
\begin{equation}
\label{eq:a14} \alpha_{\parallel}(\omega)=2
\alpha_{\perp}(\omega)=\frac{1}{4 \pi r G^{*}(\omega )}.
\end{equation}

 Similarly to the one-particle method, the measured response function must
be corrected for the trap stiffness. The trap correction for
two-particle microrheology has been explained in detail in
reference \cite{Maryam}.

\section{Results}

We carried out the measurements on a variety of Laponite
concentrations and salt contents (2.8, 3.2 wt\%, in pure water, 3
wt\% in pH=10, 1.5 wt \%, 5mM NaCl, 0.8 wt \%, 6mM NaCl, 0.8 wt
\%, 3mM NaCl ). We have chosen these samples to ensure that their rate of
aging is slow enough to guarantee that no significant aging occurs during
each measurement. On the other hand they evolve fast enough to
allow us to follow the whole evolution within a few hours. The
samples 2.8, 3.2 wt\%, in pure water, 3 wt\% in pH=10 and 1.5 wt
\%, 5mM NaCl showed the properties of a glassy sample according to
our light scattering data and samples 0.8 wt \%, 6mM NaCl, 0.8 wt
\%, 3mM NaCl behaved like a colloidal gel \cite{sarathesis,gelpaper}. In
addition, we find that a pH of 10 did not affect the aging dynamics
qualitatively. It merely acted as an electrolyte that slightly
accelerated the aging. The same held for Laponite 1.5 wt\%, 5mM NaCl.
In this case salt just accelerated the aging, but it did not
change the underlying dynamics of the aging process. As we
demonstrated elsewhere \cite{gelpaper} both samples belong to the
glass region of the phase diagram.

For the detailed discussion of our results, we focus on two samples that are
representative of the others:
one sample that behaves as a glass (C= 3.2 wt\%),
and one sample (C=0.8 wt\%, 6mM NaCl) that behaves like a gel. In the latter case,
the structure factor shows a strong q-dependence, as illustrated in Fig. \ref{fig1}.
This suggests a more heterogenous, gel-like structure, in contrast to the more
homogenous samples that we identify as a glass. In the following, we
shall characterize our samples as gels or glasses in this way.

To follow the aging of the systems we trapped a bead in a single laser trap and measured the
displacement power spectral densities (PSD) as a function of
waiting time. Since the system evolves towards a non-ergodic
state, the time average may not necessarily be equal to the
ensemble average for the measured PSDs.  However, in our range of
frequencies ($1- 10^5$ Hz) we confirmed that our results did not
depend on the time interval used to compute the time average.
Thus, we can use the time-averaged PSD without averaging over
several beads in our study.

\begin{figure}[t]
\begin{center}
\includegraphics[scale=0.60]{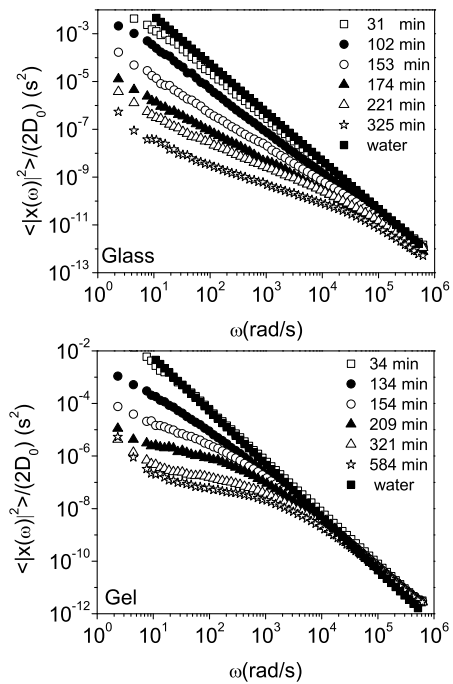}
\caption{Normalized displacement power spectral densities
$\langle |x(\omega)|^2\rangle/2D_0$ of silica probe particles in a glassy sample  (3.2
wt\%, bead diameter  1.16 $\mu$m ) and a gel-like sample (0.8 wt\%, 6mM NaCl,
bead diameter 0.5 $\mu$) in the x direction) with
increasing age after preparing the sample. Waiting times are given
in the legend. The filled squares show the PSD of a bead in pure
water for comparison. All experiments were done at
$21^{o}C$.}\label{fig:PSDs}
\end{center}
\end{figure}

 Fig.\ \ref{fig:PSDs} shows the measured displacement PSDs
as a function of frequency during the aging of the glass and the gel respectively.
We normalized the PSDs with the diffusion coefficient
$(D_0=kT/(6 \pi \eta_{water}R_{bead})$ of a same-size bead
measured in water, so that the normalized PSDs will be independent of bead
size. It is evident that in both systems the particle motion
progressively slows down with increasing aging time $t_{w}$,
reflecting the increase of viscosity in the system. The PSDs in
both samples start from a state close to water, for which
$\langle|x(\omega)|^2\rangle/2D_0=1/\omega^2$. Gradually their amplitudes as
well as the absolute values of their power-law slopes decrease with time. There is a crossover time
$t_0$ such that for $t_{w}< t_0$, the PSDs can be described by a
single power law. At longer aging times $t_{w}> t_0$, two
distinct slopes appear in the log-log plots (Fig.\ \ref{fig:PSDs}).

\begin{figure}[t]
\begin{center}
\includegraphics[scale=1.2]{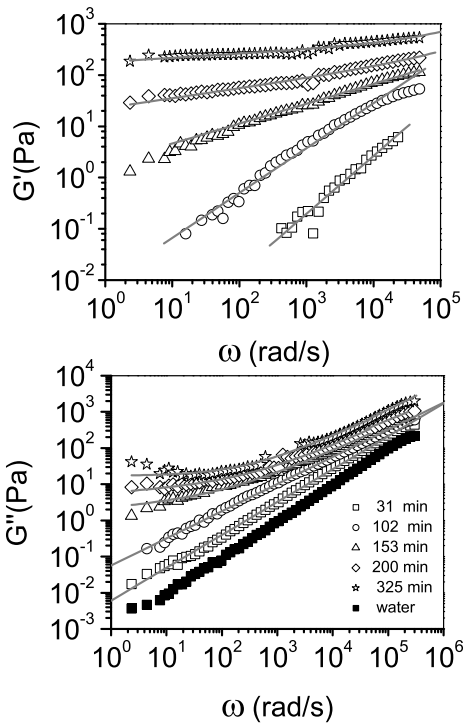}
\caption{Glass data: The symbols show the shear moduli $
G'(\omega)$ and $G''(\omega)$ (absolute magnitude) as a function of frequency measured
using 1.16 $\mu$m silica probe particles in a 3.2 wt\% Laponite
solution in pure water with increasing aging time after preparing the
sample. Aging times are given in the legend. The lines show the
fits of $ G'(\omega)$ and $G''(\omega)$ according to $C_{1} (-i
\omega) ^{a}+C_{2}(-i \omega )^{b}$ in which $C_{2}=0 $ for aging
times $t_{w} < 120 $ min. }\label{fig:Gglass}
\end{center}
\end{figure}

\begin{figure}[h!]
\begin{center}
\includegraphics[scale=0.58]{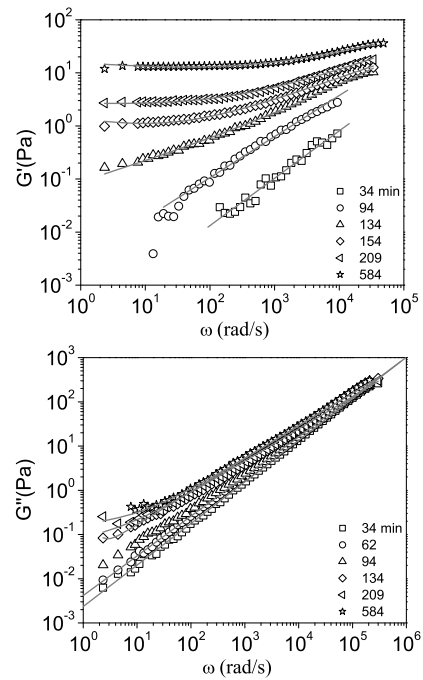}
\caption{Gel data: The symbols show the shear moduli $ G'(\omega)$
and $G''(\omega)$ (absolute magnitude) as a function of frequency measured using a 0.5
$\mu$m silica probe particle in a 0.8 wt\% Laponite solution in 6
mM NaCl water with increasing aging time after preparing the sample.
Aging times are given in the legend. The lines show the fits of
$ G'(\omega)$ and $G''(\omega)$ according to $C_{1} (-i \omega)
^{a}+C_{2}(-i \omega )^{b}$ in which $C_{2}=0 $ for aging times
$t_{w} < 100 $ min. }\label{fig:Ggel}
\end{center}
\end{figure}

The evolution of the local shear moduli $G^*$ obtained from PSDs
are shown in Fig.\ \ref{fig:Gglass} (glass) and \ref{fig:Ggel}
(gel). The shear moduli are derived from single
particle MR according to Eq.\ (\ref{eq:a9}). It is evident that the systems evolve from an initially completely viscous
to a strongly viscoelastic fluid. At the early stages of
aging, the loss modulus is still much larger than the storage modulus
$(G''\gg G')$ representing a more liquid-like state. With time the samples
become more solid-like: the elastic modulus becomes larger than
the loss modulus $(G''\ll G')$. We observe also that the changes
in $G'$ are more dramatic than the changes in $G''$. While $G''$
almost saturated after 170 min for the glass and $100$ min for the
gel, $G'$ continues to grow with time.


Visual inspection showed that the gel was ``softer" than the glass. When
we mechanically shook similar tubes containing gel or glass, the gel
liquefied at a clearly smaller stress: it appears that gels had a lower
yield stress compared to glasses. Therefore it is reasonable to expect that gels also  have a lower
viscoelastic modulus than glasses, as comparison of Fig.\
\ref{fig:Gglass} and Fig.\ \ref{fig:Ggel} indeed confirms. Furthermore,
the ratio $G'/G''$ at low frequencies is
higher in the gel ($G'/G''=30$ for Laponite 0.8 wt\%, 6mM)
compared to the glass ($G'/G''=12$ for Laponite 0.8 wt\%), at late stages of aging when
$G''$ has almost saturated and $G'$ evolves very slowly.

From microrheology we conclude that the aging behaviors of gels and glasses are
qualitatively similar. We know, however, from light scattering
measurements that the underlying structures of gels and glasses are
very different \cite{PRL}. Since spatial heterogeneity is the defining feature of gels, we thus set out to investigate if
local measurements of microrheology across the samples can detect the
difference between gels and glasses.

\subsection{Heterogeneity}

Heterogeneities within a sample can be explored by measuring the
PSDs of multiple beads at different positions in the sample.
 A discrepancy between the shear moduli obtained from  one- and two-particle MR can also
be used as an indicator of a heterogeneous structure.  A further test
of heterogeneity in a material is provided by comparison of MR
with bulk rheology, as will be discussed below. To
 investigate the homogeneity of colloidal
 gels and glasses of Laponite, we performed two types of measurement.
 First, we made simultaneous measurements of PSDs of two independent
beads in two independent traps at different stages of aging. In
another set of experiments, we measured PSDs of multiple beads in
 aged gels and glasses. The results of our experiments for both
 gels and glasses will be discussed below.

\subsubsection{Glass}

\begin{figure}[t]
\begin{center}
\includegraphics[scale=1.3]{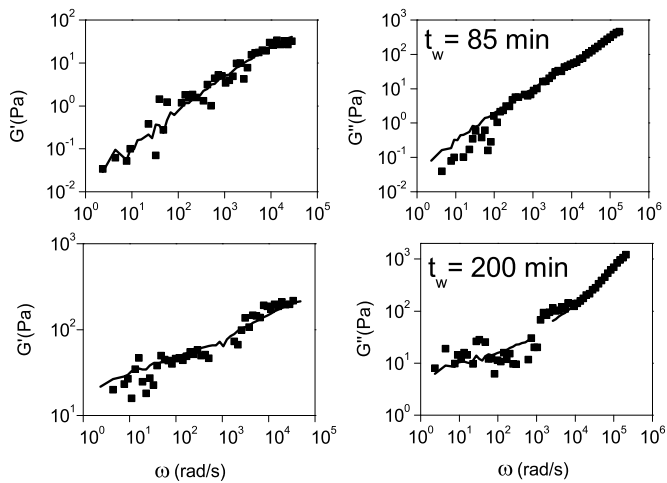}
\caption{ Glass data: The shear moduli $ G'(\omega)$ and
$G''(\omega)$ (magnitude) at two different stages of aging in a 3.2 wt\%
Laponite solution derived from one- (lines) and two-particle (symbols) MR using
1.16 $\mu$m silica probe particles. The distance between the two
particles was 6 $\mu$m. The aging times are shown in the figure.
Note that in the late stages of aging the material becomes too stiff to
 obtain a good cross-correlation signal between the two beads over the background noise. }\label{fig:2PMRglass1}
\end{center}
\end{figure}

For the glassy samples the displacement PSDs turned out to be
independent of the bead position, as was concluded from a
comparison of simultaneous measurements of PSDs of two independent
beads in two independent traps during aging. Furthermore, the
comparison between one- and two-particle MR reveals that within
the experimental error,
 the complex shear moduli are identical to within the experimental error between the two methods for all stages of aging as
  shown in Fig.\ \ref{fig:2PMRglass1}.

\begin{figure}[h!]
\begin{center}
\includegraphics[scale=0.6]{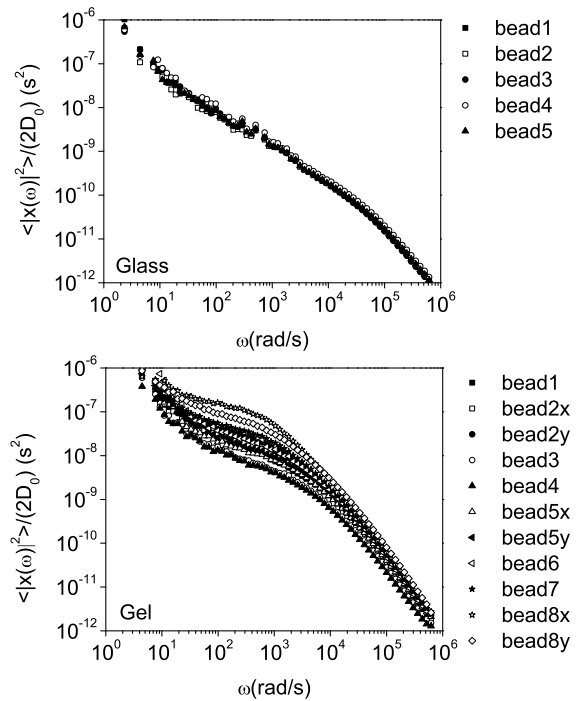}
\caption{The displacement PSDs of 0.5 $\mu$m silica beads
measured at different positions within an aged glass (Laponite 3.2 wt
\% in pure water, $t_w\approx5$ h ) and within an aged colloidal gel
(Laponite 0.8 wt\% in 6mM NaCl solution, $t_w\approx10$ h ).
}\label{fig:inhomgenous}
\end{center}
\end{figure}

\begin{figure}[t]
\includegraphics[scale=0.6]{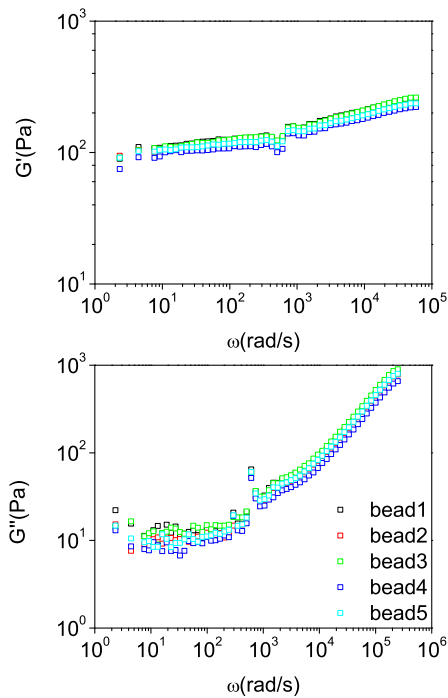}
\caption{Local elastic modulus $G'$ and loss modulus $G''$
measured at different positions in an aged glassy sample of
Laponite 3.2 wt \% in pure water ($t_w\approx5$h)
}\label{fig:hetroglass}
\end{figure}

This was further verified by measuring the PSDs of several beads
at different positions of an aged sample. As can be seen in Fig.\
\ref{fig:hetroglass}, the measured shear moduli
were independent of the position of the bead in the sample, verifying
the homogeneity of the glassy sample, as shown also in Fig.\
\ref{fig:inhomgenous}a.

These results suggest that the Laponite glass has a homogenous
viscoelasticity, at least on length scales larger than half a micrometer, which is the length scale one-particle MR intrinsically averages over. An additional check on this can be obtained
from a comparison between microrheology and macrorheology which
should yield the same results if the sample is homogeneous.

\begin{figure}[h!]
\begin{center}
\includegraphics[scale=0.6]{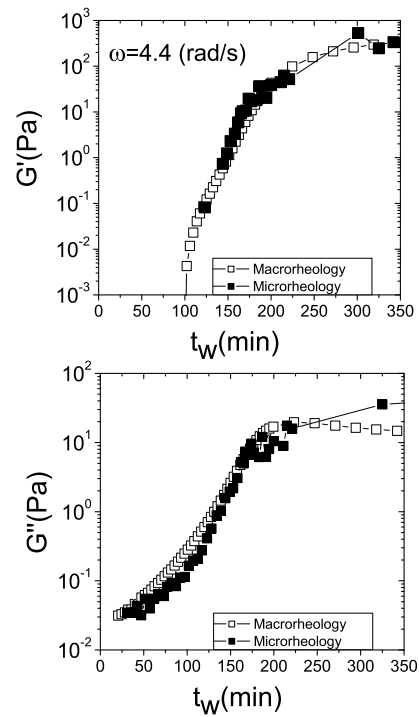}
\caption{ Elastic and loss modulus as a function of aging
time for a sample of Laponite 3.2 wt \% in pure water obtained
from macrorheology and one-particle MR at $f=0.7$ Hz. The strain amplitude in the macrorheology measurements was 0.01.}\label{fig:rheology}
\end{center}
\end{figure}

Figure \ref{fig:rheology} shows the shear moduli extracted from
MR and macrorheology experiments at a fixed frequency of
($f = 0.7$ Hz) during the course of aging. The overall agreement between macrorheology and MR
is good. For the early stages of aging, the $G''$ measured by the
macrorheometer appears slightly higher, but this can be attributed to
the large moment of inertia of the rheometer bob; macrorheology
does not provide accurate measurements of the shear moduli when
$G^{*} < 1$ Pa. MR, on the other hand, has other sources
of errors at low frequencies, especially for the late stages of
aging, when the material becomes very rigid. In this case, the
signal detected by the photodiode becomes small compared to the noise level; $1/f$ laser pointing noise
dominates at low frequencies. This is the most plausible
explanation for the slight discrepancy between the results from the two methods at
long aging times.

\subsubsection{Gel}

 Measuring the displacement PSDs of several beads at different positions of a gel at the late
stages of aging revealed a considerable degree of inhomogeneity
(Fig.\ \ref{fig:inhomgenous}). Not only were the PSDs position
dependent, but at some positions in the sample the measured
PSDs were also anisotropic, i.e. fluctuations in x and in y direction gave different results.

\begin{figure}[h!]
\begin{center}
\includegraphics[scale=0.6]{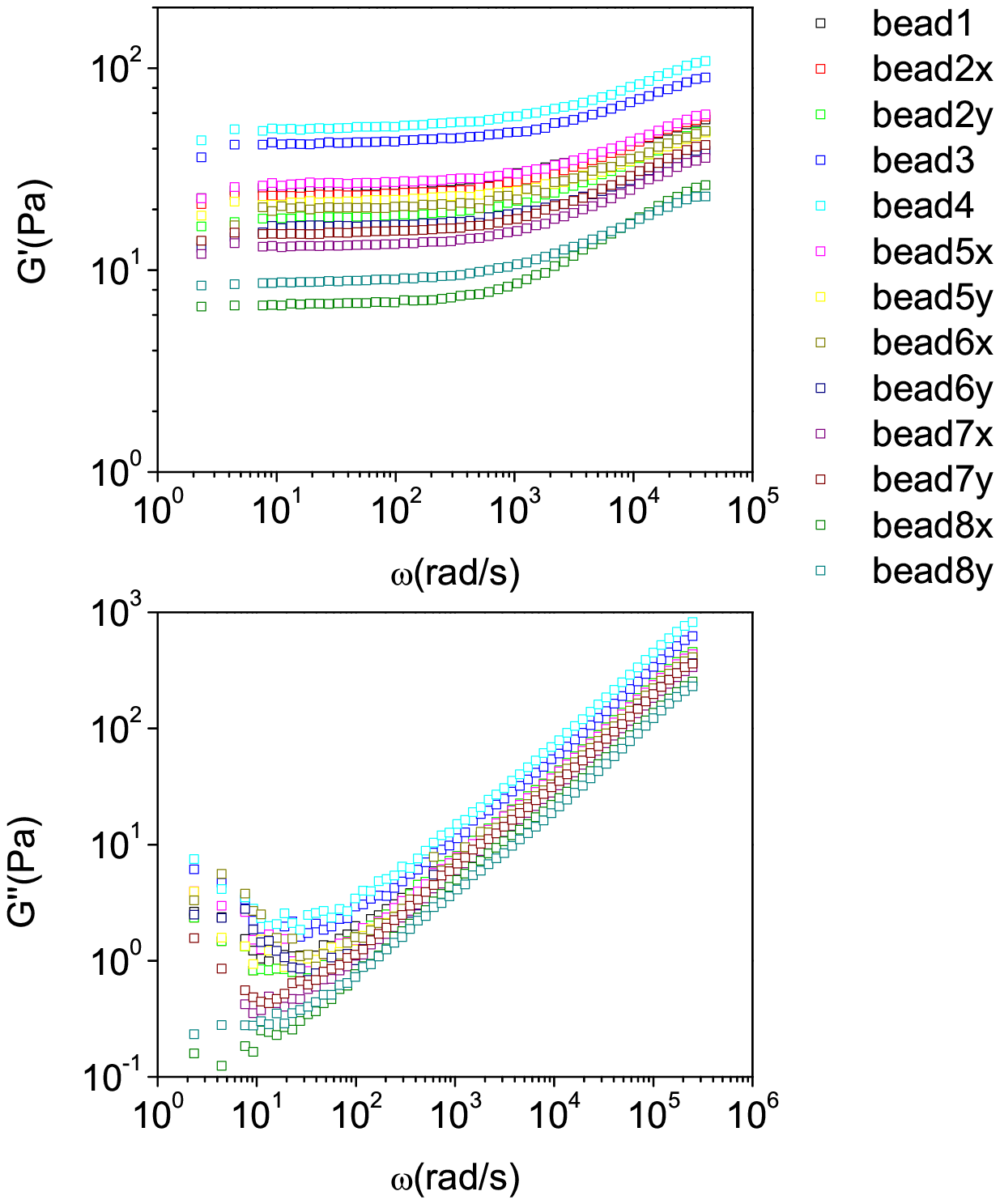}
\caption{Gel data: local elastic modulus $G'$ and loss modulus
$G''$ measured at different positions in an aged gel sample of
Laponite 0.8 wt \% in 6mM NaCl solution ($t_w\approx10$h)
}\label{fig:hetrogel}
\end{center}
\end{figure}

This result is consistent with the static light scattering measurements
for this sample shown in Fig.\ \ref{fig1} that suggest inhomogeneities of the gel on a length scale comparable to the inverse scattering vector, i.e. micrometers. Therefore, exploring such a gel using microrheology
with a probe on the order of the mesh size of the network, one can detect these characteristic inhomogeneities. In (Fig.\ \ref{fig:hetrogel}) we have plotted the shear moduli seen by the
beads at different positions. It can be
seen that there was an order of magnitude difference between the
smallest and largest elastic moduli measured in the same sample and at the same time.

\begin{figure}[h!]
\begin{center}
\includegraphics[scale=0.9]{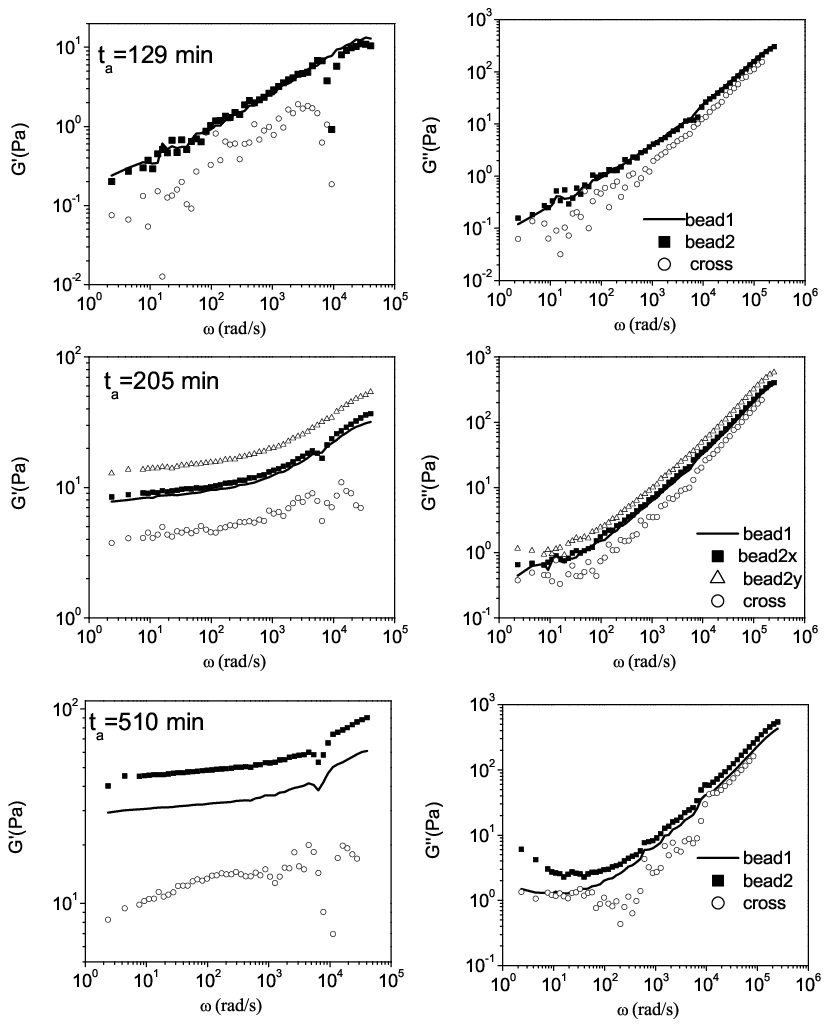}
\caption{Gel data: The shear moduli $ G'(\omega)$ and
$G''(\omega)$ at different stages of aging derived from one- and
two-particle MR of 0.5 $\mu$m silica probe particles in a gel of
Laponite 0.8 wt \% in 6mM NaCl solution. The distance between the
two beads was 4.66 $\mu$m. For $t_a=129$ min the shear moduli at two different positions are equal.  At  $t_a=205$ min the shear moduli at the two positions are not equal. Furthermore shear modulus at position 2 shows anisotropy. At a later time  $t_a=510$ min the shear moduli at the two positions are not equal but the anisotropy observed earlier at position of bead 2 has been disappeared.
}\label{fig:2PMRgel1}
\end{center}
\end{figure}

It is intriguing to ask when the heterogeneity starts to develop in the aging samples. It is likely to appears as a
network-like structure is building up in the gel. To answer this
question, we measured the PSDs of two beads at different positions
of a gel as a function of aging time. We performed two sets of
experiments: in the first one the two beads were positioned at a
relatively close distance $r=$4.66 $\mu$m (Fig.\ \ref{fig:2PMRgel1})
and in the other one at a large distance of $r=19$  $\mu$m.

In both experiments, the responses, i.e. the PSDs at
different positions were equal in the early stages of aging. However, as time progressed,
the PSDs measured at
different positions began to differ. In addition, at
later stages of aging, the displacement PSDs measured for some of
the beads became anisotropic, meaning that the PSDs in the $x$ and
$y$ directions were not equal anymore. In some measurements the
anisotropy survived the latest measurement. For some other
measurements, the anisotropy disappeared after some time (look at
Fig.\ \ref{fig:2PMRgel1} for example). This suggests that the
building up of structure in the gel is a dynamic process; at some
points and times  more particles join to the network and at some
other points and times some particles disintegrate from the
network.

Furthermore, our experiments showed that immediately after
preparation, shear moduli obtained from two-particle MR and
one-particle MR were equal. But already at relatively early stages of
aging, the two-particle MR results differed from 1PMR results as demonstrated in
Fig.\ \ref{fig:2PMRgel1}. This deviation appeared long before the
local shear moduli of the two beads in one-particle MR started to differ. For more details, see also \cite{sarathesis}.

\begin{figure}[t]
\includegraphics[scale=0.6]{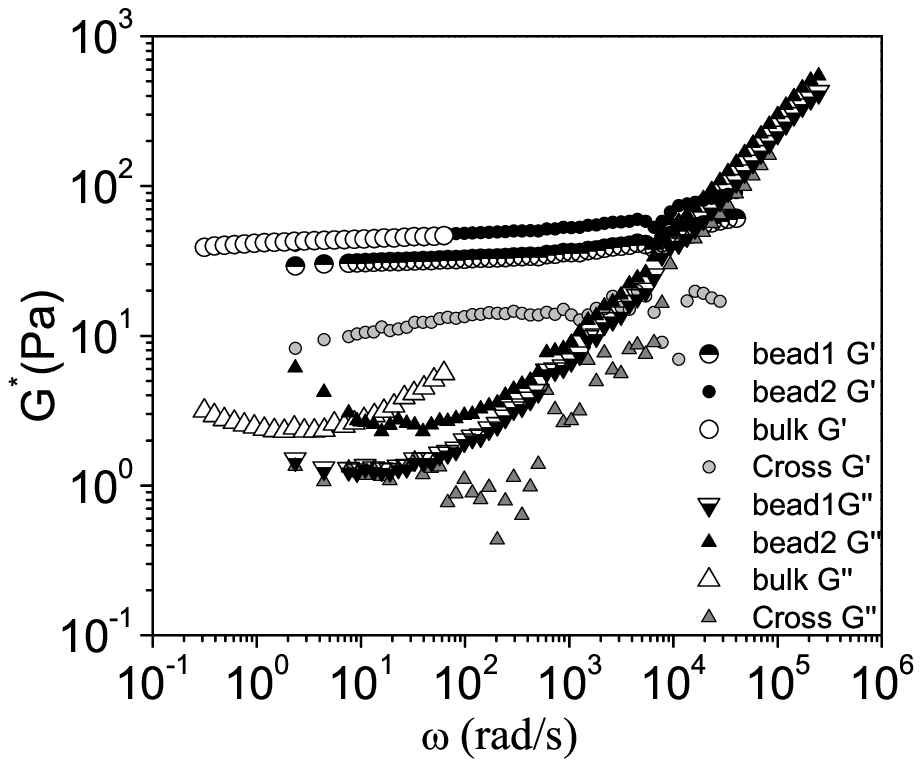}
\caption{Complex shear modulus at a late stage of aging
$t_w\approx8.5$h obtained from single-particle MR at two different positions of the sample,
two-particle MR and bulk rheology in a sample of Laponite 0.8
wt\%, 6mM NaCl. The circles show $G'$ and triangles show $G''$
values. }\label{gelbulk}
\end{figure}

Our measurements on several bead pairs at varying distances suggest
that these inhomogeneities extend over a range of at least 100 micrometers.
Therefore the macroscopic bulk shear modulus is not necessarily expected to be
equal to that measured by single particle MR.  In Fig.\
\ref{gelbulk}, we compare the shear moduli obtained from  one- and
two-particle MR with the results of macrorheology at late stages
of aging ($t_w\approx 8.5$ h) when the changes in the loss and elastic
moduli are slow. It is evident that the local shear modulus measured at one of
the positions in the sample was equal to the bulk value, while the
others reported a considerably lower shear modulus. Notably, the shear
modulus obtained from the cross correlation of two-particles is
lower than both bulk and local shear moduli. This suggests that
two-particle MR can be used to detect inhomogeneities as long as they occur on length scales below the distance between the particles, but the results may still not reflect bulk properties if heterogeneities extend beyond the scale of the inter-particle distance.

\subsection{Model for the viscoelastic behavior}
\begin{figure}[h!]
\begin{center}
\includegraphics[scale=0.85]{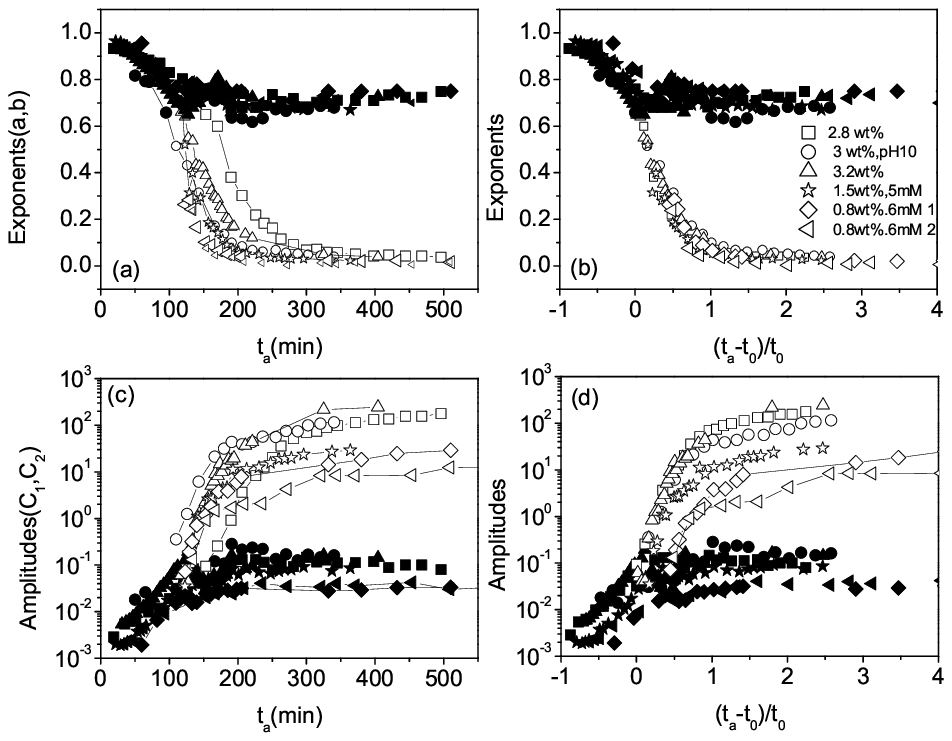}
\caption{The complex shear moduli of Laponite suspensions can be
described as the sum of two power laws $C_{1} (-i \omega)
^{a}+C_{2}(-i \omega )^{b}$ in which $C_{2}=0 $ for waiting times
$t_{w} < t_{0}$. The crossover times are $t_{0}= 155, 95, 120,
105,95 $ min  for Laponite concentrations 2.8 wt\%, 3 wt\%,pH=10,
3.2 wt\% , 1.5 wt\%, 5mM NaCl and two different positions of 0.8 wt\%, 6mM NaCl,
respectively. (a) The evolution of power-law exponents $a$ (filled symbols)
and $b$ (open symbols) as a function of aging time for
different concentrations of Laponite. (b) The exponents $a$
(filled symbols) and $b$ (empty symbols) as a function of scaled
aging time (c) The amplitude of viscoelastic contributions
$C_{1}$ (filled symbols) and $C_{2}$ (open symbols) as a function
of aging for different samples. (d) The same as panel (c) but plotted
versus scaled aging time.
 The sample concentrations are shown in the legend.
}\label{fig:fitslap}
\end{center}
\end{figure}

It has been noted in the context of weakly attractive colloids
\cite{Trappe} and biopolymer networks \cite{gardel2004} that the
addition of two power law contributions describes the shear modulus
very well. This result appears to reflect the existence of two
distinct contributions to the viscoelasticity of the system and
can be interpreted as a superposition of a more elastically rigid network
(weakly frequency-dependent) and viscoelastic background (with a
strong frequency dependence).

Our data ( Fig.\ \ref{fig:Gglass} and  Fig.\ \ref{fig:Ggel}) can be interpreted in a similar manner:
In addition to a strongly frequency-dependent viscoelastic response at high
frequencies, a more elastic (weakly
frequency-dependent) response appears after some aging time $t_0$ and slowly increases in amplitude during the aging process.
To be more
precise, we see that the complex shear modulus of both gels and
glasses crosses over from a single power law to a superposition of
two power laws around a certain waiting time $t_0$ which depends on
the sample $(t_{0}\approx 155 min $ for the glass sample of
Laponite 3.2 wt \% and $t_{0}\approx 95 min $ for the gel sample
of Laponite 0.8 wt \%,6mM NaCl )  \cite{jabbari}. The local shear
moduli of both samples turn out to be well-described by the
following expression:
\begin{eqnarray} \label{eq:twopower}
G(\omega)&=& G'(\omega)+ i G''(\omega) \nonumber \\
 &\equiv& \left\{\begin{array} {l@{\quad: \quad}l}
 C_{1} (-i \omega) ^{a}& t_{w} < t_{0} \\ C_{1} (-i \omega) ^{a}+C_{2}(-i \omega )^{b}& t_{w} > t_{0}
 \end{array} \right.
   \end{eqnarray}
Physically, this model implies that to two distinct stresses
arise under a common imposed strain --- in the way forces add for springs in parallel,
as opposed to displacements/compliances that would add for springs in series.
This response would be expected, for instance, for two interpenetrating structures/systems
that displace together under strain, at least on the scale of our probe particles,
which are large compared to the individual Laponite particles.
This can, in principle, be the case whether or not the material appears to be
homogeneous on this scale. A tenuous elastic network structure immersed in a more
fluid-like background, such as we might expect for a gel,
would behave in this way, provided that the network and the
background medium are strongly coupled hydrodynamically, and that the network
spans length scales corresponding to the imposed strain. Such a gel could
appear to be either homogenous or heterogeneous, depending on the length scale probed.

We find that the exponent of the single power law decreased from 1 to a value
of about 0.7 before the second component becomes visible.
The exponent and amplitude of the first component  $C_{1} (-i \omega) ^{a}$ do not change further
with aging time for $t_{w} > t_{0}$ while the amplitude of the
other one  $C_{1}$ grows appreciably over the same times.

In Fig.\ \ref{fig:fitslap}(a) and (c), we have plotted
the evolution of the fitting parameters as a function of aging
time for different samples. As can be seen in the figure, the development of the
two viscoelastic components for different samples is
qualitatively similar, although the rate of change depends on sample
concentration and salt content.

Interestingly, the evolution curves of the exponents $a$ and $b$
for the different samples superimpose if we scale the aging time as
$t'_{a}=(t_{w}-t_{0})/ t_{0}$. The crossover times are $t_{0}=
155, 95, 120, 105, 95 $ min  for Laponite concentrations 2.8 wt\%,
3 wt\%, pH10, 3.2 wt\%, 1.5 wt\%, 5mM  and 0.8 wt\%, 6mM NaCl,
respectively. For the amplitudes on the other hand, the data do
not collapse. Especially the amplitudes of the second
(viscoelastic) component systematically decreases as the Laponite
content is reduced. Furthermore, for the gel the amplitude depends on the position and we can see some
fluctuations in the amplitude of the second component $C_2$, at
later stages of evolution. This can be understood in terms of the
dynamic process of gel formation in which Laponite particles still
can join or detach from the network.

\section{Discussion and Conclusion}

We have studied the evolution of the viscoelastic properties of a
variety of Laponite suspensions including both gel-like and glassy states over
a wide range of frequencies using macro- and micro-rheology
techniques. Our measurements reveal the differences between the
mechanical properties of gels and glasses.

 The glassy samples are homogenous on all length scales probed in our experiments ($l>0.5
\mu$m). This is further confirmed by comparing microrheology and
conventional macrorheology results. We find that measurements at
different scales all give the same results. Thus, there is no
evidence for spatial inhomogeneity as expected for glassy
systems in general.

 In the gels, however, along with the evolution from a liquid-like state to a viscoelastic state, inhomogeneities develop in
time.  These inhomogeneities are detected by measuring the local shear moduli at
different positions within the samples at nearly equal waiting times.

\begin{figure}[t]
\includegraphics[scale=0.6]{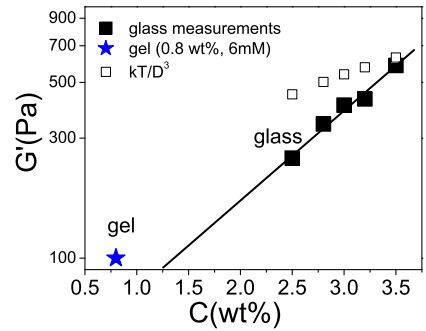}
\caption{Bold symbols: The elastic modulus obtained from
macrorheology at late stages of aging at the aging time that there is a
crossover from a fast regime of aging to a slower regime as a
function of concentration measured for different glassy samples
and a gel sample of 0.8 wt\%, with 6mM salt at $f=0.05$ Hz
. Empty symbols: An estimate of the plateau
value $G'_{P}\propto kT/D^3$ is shown for comparison.}\label{GC}
\end{figure}

Another difference between gels and glasses that we observed is that the
change in the ratio $G'/G''$ is much more rapid for a gel than for a
glass as the material evolves from a purely viscous liquid to
solid-like system.

When we track the values of the elastic modulus at a fixed frequency (here 0.05 Hz) at late stages of aging,
when there is a crossover from a fast aging rate to a slower aging
rate, we find that the elastic
modulus scales linearly with concentration. One can roughly estimate the plateau value for glasses as $G'_{P}\propto kT/D^3$, where $D$
is the characteristic structural length of the system. We have
taken $D$ as half of the interparticle distance. This estimate
predicts the order of magnitude fairly well. As shown in Fig.\
\ref{GC}, the extrapolation of our data suggests that no glassy
samples exist at concentrations lower than 1.6 wt\%.

Therefore the elasticity for samples with lower concentrations should
stem from a different mechanism. Indeed, in such samples the aging
proceeds through gel formation. For comparison, we have shown the
elastic modulus of a gel sample of 0.8 wt\%, 6mM in Fig.\
\ref{GC}.

Despite the differences between gels and glasses , we find a similar
frequency dependence of the visco-elastic moduli
for gels and glasses. The local viscoelastic moduli for both gels
and glasses cross over from a single
power law to the sum of two power laws around a certain time $t_0$. These results demonstrate
the existence of two distinct contributions in the viscoelasticity
of the system in the later stages of aging. In addition to a
strongly frequency-dependent viscoelastic shear modulus at high
frequencies $\cong \omega^{0.7}$, we also observe the slow
development of a more elastic (only weakly frequency-dependent)
shear modulus during the aging. The exponents of the power laws
follow exactly the same time course of evolution for different
concentrations if we scale the aging time as
$t'_{a}=(t_{w}-t_{0})/ t_{0}$. This result is independent
of the sample being a gel or a glass.

 The crossover from a single
frequency-dependent component to a superposition of a strongly
frequency-dependent viscoelastic component plus a weakly frequency
dependent (elastic) component was previously interpreted in the
context of polymer networks as being due to large inhomogeneities
\cite{deGennes,gardel2004}. Here the sum of two power-laws
describes both gel (heterogenous) and glass (homogenous) local
shear moduli, suggesting that locally the underlying physical
process responsible for the evolution of gels and glasses is
similar. This poses the rather puzzling question what the physical
origin of the two power-laws in the viscoelasticity is.

 \textbf{Acknowledgments} This research has been supported
by the Foundation for Fundamental Research on Matter (FOM), which
is financially supported by Netherlands Organization for
Scientific Research (NWO). LPS de l'ENS is UMR8550 of the CNRS,
associated with the universities Paris 6 and 7. C.F.S was further supported by the DFG Center for the Molecular Physiology of the Brain (CMPB).

\end {document}